\documentclass[twocolumn,english,aps,english,preprintnumbers,groupedaddress,amsmath,amssymb,prl]{revtex4}
\usepackage[T1]{fontenc}
\usepackage[latin9]{inputenc}
\usepackage{amsmath}
\usepackage{amssymb}
\usepackage{esint}

\makeatletter
\@ifundefined{textcolor}{}
{%
 \definecolor{BLACK}{gray}{0}
 \definecolor{WHITE}{gray}{1}
 \definecolor{RED}{rgb}{1,0,0}
 \definecolor{GREEN}{rgb}{0,1,0}
 \definecolor{BLUE}{rgb}{0,0,1}
 \definecolor{CYAN}{cmyk}{1,0,0,0}
 \definecolor{MAGENTA}{cmyk}{0,1,0,0}
 \definecolor{YELLOW}{cmyk}{0,0,1,0}
}

\makeatother

\usepackage{babel}
\begin{document}
\title{Comment on ``Inverse Doppler shift and control field as coherence

generators for the stability in superluminal light''}
\author{Bruno Macke}
\author{Bernard S\'{e}gard}
\email{bernard.segard@univ-lille.fr}

\affiliation{Universit\'{e} de Lille, CNRS, UMR 8523, Physique des Lasers, Atomes et
Mol\'{e}cules, F-59000 Lille, France}
\date{April 25, 2019}
\begin{abstract}
In their study of inverse Doppler shift and superluminal light {[}Phys.
Rev. A \textbf{91}, 053807 (2015){]}, Ghafoor \emph{et al}. consider
a three-level atomic arrangement with transitions in the optical domain.
In fact, the values they give to the parameters lead to a probe wavelength
lying in the decimeter band. We point out that the Doppler shifts
are then negligible and remark that the simulations performed by Ghafoor
\emph{et al.} do not evidence any superluminal effect. 
\end{abstract}
\maketitle
In their study of inverse Doppler shift and superluminal light \cite{art1},
Ghafoor et al. consider a three-level atomic arrangement with transitions
in the optical domain, referring in particular to the sodium $D_{2}$
line at $586.9\,\mathrm{nm}$. On the other hand, they specify in
the caption of their Fig.2 that all the (angular) frequencies are
given in units of $\Gamma=2\pi\times1\,\mathrm{MHz}$ and that the
frequency of the probe transition $\omega_{ac}=1000\Gamma$. The corresponding
wavelength is thus $\lambda=30\,\mathrm{cm}$ (in the decimeter band).

The first consequence of the large value of the probe wavelength is
that the Doppler broadening $V_{D}$ is very small, in the order of
$2\pi\times1\,\mathrm{kHz}$ . The consideration of $V_{D}$ going
from $2$ to $12\,\mathrm{MHz}$ as made in \cite{art1} is meaningless
and the so-called inverse Doppler shift, claimed as the novelty of
the article, is in fact negligible.

A second point is that the calculations developed in \cite{art1}
lead to fully unrealistic values of the atomic number density $N$.
As correctly given in the article, the electric susceptibility for
the probe reads in SI units: 
\begin{equation}
\chi=\frac{2N\left|\wp_{ac}\right|^{2}\rho_{ac}}{\varepsilon_{0}\hbar\Omega_{p}}\label{eq:un}
\end{equation}
where $a$ ($c$) is the upper (lower) level of the probe transition,
$\wp_{ac}$ ($\rho_{ac}$) is the corresponding matrix element of
the dipole moment (of the density operator) and $\Omega_{p}$ is the
Rabi (angular) frequency of the probe. From the involved discussion
following this equation, it results that 
\begin{equation}
\Gamma=\frac{\left|\wp_{ac}\right|^{2}\omega_{ac}^{3}}{\varepsilon_{0}\hbar c^{3}}=O\left(\frac{N\left|\wp_{ac}\right|^{2}}{\varepsilon_{0}\hbar}\right)\label{eq:deux}
\end{equation}
and that 
\begin{equation}
N=O\left(\frac{8\pi^{3}}{\lambda^{3}}\right)\label{eq:trois}
\end{equation}
For $\lambda=30\,\mathrm{cm}$, we get an atomic number density in
the order of $10^{-2}\mathrm{cm^{-3}}$, which is 12 orders of magnitude
lower than those attainable in the best vacuum devices.

As a third point, we remark that, contrary to the claim made in the
article title, the simulations made in \cite{art1} do not evidence
any superluminal effect, namely an advance of the intensity profile
of the transmitted pulse on that of the incident one (see Fig.5).
The calculation itself raises some questions. The transmitted field
is actually the inverse Fourier transform of $S_{in}(\omega)H(\omega)$,
where $S_{in}(\omega)$ and $H(\omega)$ are, respectively, the Fourier
transform of the incident field and the transfer function of the medium.
Insofar as $S_{in}(\omega)$ is Gaussian and $H(\omega)$ is the exponential
of a polynomial of degree 3, the result cannot be that given by Eq.(15)
in \cite{art1} but necessarily involves an Airy function. We also
note that the transfer function $H(\omega)$ considered by Ghafoor \emph{et al}. neglects the frequency dependence of the medium transmission
that can considerably affect the profile of the transmitted pulse
\cite{art2}.

For completeness, we mention that some equations in \cite{art1} seem
to be dimensionally inhomogeneous, that the Einstein's coefficient
given below Eq.(4) is erroneous (see \cite{art3} for its exact value
in SI units) and that Eq.(1) and Eq.(8) mix results that hold, respectively,
in SI and in electrostatic units (the corresponding susceptibilities
differ by a factor of $4\pi$).

We finally point out that the atomic number density given by Eq.(\ref{eq:trois}),
anomalously weak in the conditions considered in \cite{art1}, raises
on the contrary to values $N=O(10^{15}\mathrm{cm^{-3}})$ which are
too large when the probe wavelength $\lambda$ is that of the sodium
$D_{2}$ line. On another hand, the fixed ratio $\omega_{ac}/\Gamma=1000$
leads then to lifetimes of the excited atomic states which are fully
unrealistic (in the subpicosecond range).

We thank Shubhrangshu Dasgupta from the Indian Institute of Technology
Ropar for helpful information. This work has been partially supported
by the Minist\`{e}re de l'Enseignement Sup\'{e}rieur, de la Recherche et de
l'Innovation , the Conseil R\'{e}gional des Hauts de France and the European
Regional Development Fund (ERDF) through the Contrat de Projets \'{E}tat-R\'{e}gion
(CPER) 2015--2020, as well as by the Agence Nationale de la Recherche
through the LABEX CEMPI project (ANR-11-LABX-0007).

\end{document}